\documentclass[12pt,twoside]{article}
\usepackage{fleqn,espcrc1}
\usepackage{amsmath,amssymb,epsfig,wrapfig}

\title{$J/\psi$ suppression in central Pb--Pb collisions}

\author{P. M. Dinh\address[SPhT]{Service de physique th\'eorique,
        CEA-Saclay, F-91191 Gif-sur-Yvette cedex},
        J.-P. Blaizot\addressmark[SPhT]\thanks{Member of CNRS},
        and
        J.-Y. Ollitrault\addressmark[SPhT]$^*$}
       
\begin{document}

\maketitle

\begin{abstract}
We discuss the recent NA50 $J/\psi$ production data in Pb--Pb
collisions, in particular the second drop at high transverse 
energies which correspond
 to the most central collisions. Using a 
model which relates the $J/\psi$ suppression to the local energy
density, we show that the data can be explained by taking into
account transverse energy fluctuations at a given impact
parameter. Predictions of this model for RHIC are briefly discussed.
\end{abstract}

\section{Introduction}

The rate of $J/\psi$ production in $p$--$p$, $p$--$A$
and $A$--$B$ collisions involving oxygen and sulphur projectiles
is well understood in terms of a hard production of the $c\bar c$ pair
followed by nuclear absorption \cite{smallsyst,Kharzeev:1997yx}. 
In Pb--Pb collisions, however, 
evidence for an additional suppression mechanism, the 
so-called anomalous suppression, has been obtained by the 
NA50 collaboration \cite{Gonin:1996wn}. 
Furthermore, in the recent NA50 data \cite{Abreu:2000ni} 
a second drop in the pattern of the $J/\psi$ production occurs 
at high transverse energy ($E_T$), that is, for the most central
collisions. The origin of this second drop is the focus of this
contribution. 

\section{Improved geometrical model} 

\begin{wrapfigure}[35]{b}{8.7cm}
\begin{center}
\epsfig{file=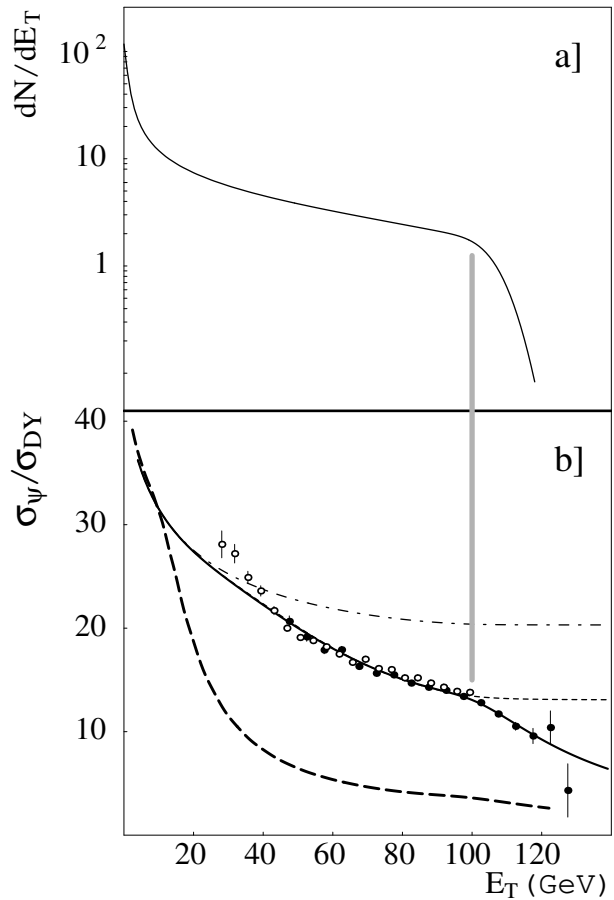,height=10.8cm}
\caption{a) Minimum bias $E_T$--distribution.  b) $J/\psi$ over
  Drell--Yan production ratio as a function of  $E_T$. 
  Open (resp. closed) circles: NA50 1996 (resp. 1998) data. 
  Dotted-dashed curve: nuclear absorption alone. 
  Full curve (resp. short dashes): our model for anomalous suppression 
  with (resp. without) $E_T$ fluctuations. Long dashes: 
  prediction for Au--Au  collisions at RHIC energies (see text).
  The vertical line sets the position of the knee of the 
  $E_T$--distribution, where the second drop occurs in the 
  ratio.}

\label{f:plot}
\end{center}
\end{wrapfigure}

It is possible to explain the anomalous $J/\psi$ suppression by using 
a scenario \cite{Blaizot:1996nq} which relates 
it to the local energy density $\epsilon$. More precisely, 
we formulate a simple geometrical model in which final state 
interactions suppress all the $J/\psi$'s originating from $c\bar{c}$
pairs produced in a region where the local energy density $\epsilon$
exceeds some critical threshold $\epsilon_c$. 
We then assume that the local energy density is  proportional to 
the density $n_p$ of participant nucleons per unit area (transverse to 
the collision axis) calculated in a 
Glauber model \cite{Glauber}: the total transverse energy is 
then proportional to the number of participants, as observed 
experimentally.
In such a model, the suppression criterion depends only 
on the impact parameter (through $n_p$), so that the $J/\psi$
suppression saturates at high $E_T$ where the impact parameter is
essentially zero (see the short-dashed curve in Fig.\ref{f:plot}b). 
A clear deviation from this simple behavior is seen in the most recent
NA50 data.

A better description is obtained by taking into account 
the fluctuations of $E_T$ for a given 
impact parameter $b$ \cite{Kharzeev:1997yx}.
These fluctuations are an essential component of the tail of the 
$E_T$ distribution (see Fig.\ref{f:plot}a). 
As a simple ansatz, we take the distribution of $E_T$ at fixed $b$ 
to be a gaussian 
with mean value $\langle E_T \rangle (\mathbf b)
= q N_p(\mathbf b)$, where
$N_p(b)=\int \textrm d^2\mathbf s \ n_p(\mathbf
s,\mathbf b)$ is the total number of participants at impact parameter
$\mathbf b$ and $n_p(\mathbf s,\mathbf b)$ the corresponding 
density per unit area at the transverse coordinate $\mathbf s$. 
The dispersion of the gaussian is given by $\sigma^2_{E_T}=a q^2
N_p(\mathbf b)$, with $a$ a dimensionless parameter. 
The values of the fit parameters $q=0.274$~GeV and $a=1.27$ 
are those determined by the NA50 collaboration \cite{private}. 

Part of the $E_T$ fluctuations is physical, part is due 
to the intrinsic resolution of the NA50 electromagnetic calorimeter. 
The latter is given by 
$\sigma_{\rm intrinsic}(E_T)/E_T=\alpha/ \sqrt{E_T}+\beta$, with 
$\alpha\simeq 0.2~{\rm GeV}^{1/2}$ and $\beta\simeq 0.005$
\cite{thesechevrot}. 
The physical $E_T$ fluctuations 
$\sigma_{\rm phys}=\sqrt{\sigma_{E_T}^2-\sigma_{\rm intrinsic}^2}$
\cite{private}
thus differ only by 10\% from the observed fluctuations
$\sigma_{E_T}$ for central collisions ($E_T\sim 100$~GeV), 
so that we neglect the intrinsic resolution in what follows.

In order to take into account the $E_T$ fluctuations, we replace the
previous estimate $\epsilon\propto n(\mathbf s,\mathbf b)$
\cite{Blaizot:1996nq}, with the more accurate $\epsilon \propto
(E_T/\langle E_T \rangle (\mathbf b)) n_p(\mathbf s,\mathbf b)$
\cite{Blaizot:2000ev}. Thus for a given impact parameter, the energy
density is proportional to $E_T$ \cite{Capella:2000zp}. 
With this prescription, the average energy density in the 
collision area $S$, defined by 
$\langle\epsilon\rangle\equiv (1/S)
\int_S \epsilon(\mathbf s,\mathbf b)d^2\mathbf s$, 
is proportional to $E_T/S$, in agreement with 
the traditional Bjorken estimate \cite{Bjorken:1983qr}.

\section{Comparison with NA50 data}

It is easy to understand that this improvement leads to 
an increased suppression at high $E_T$. Indeed, at high $E_T$ where
the geometry of the collision is essentially fixed at zero impact
parameter, the local energy density $\epsilon$ scales like 
$E_T$ and the region where it exceeds
$\epsilon_c$ becomes bigger and bigger as $E_T$ increases. 

This model reproduces the main features of the 
$J/\psi$ production pattern observed by 
NA50; however, it does not provide a perfect 
fit to the data \cite{Blaizot:2000ev}. 
A better description is obtained if we allow 
for two thresholds, in line with the idea of successive meltings 
of the $\chi$ and the $J/\psi$ (full curve in Fig.\ref{f:plot}b). 
However we should emphasize that an identically good fit is obtained
for a gradual suppression above a single threshold. Therefore the
structure in the pattern of $J/\psi$ suppression cannot be
interpreted as a signal of the successive meltings of charmonium
resonances. In any case, we stress that the convolution of the
suppression factor with the impact parameter distribution (for a given
$E_T$) and the energy density profile in the interaction region tends
to smoothen any threshold, even that associated with the onset of 
the anomalous suppression at lower $E_T$. 

Whatever the chosen scenario and/or the value of the 
thresholds, we always obtain a second drop
starting around the knee of the $E_T$--distribution, which directly
reflects the effect of $E_T$ fluctuations. 
Note that in order to account for the data, all the $J/\psi$'s must be 
suppressed at the highest energy densities. 

We have estimated the average transverse momentum squared 
of the produced $J/\psi$'s as a function of $E_T$. 
Analogous computations have been done in \cite{Kharzeev:1997ry} 
and we reproduce the behavior predicted there:
at low $E_T$, $\langle p_T^2\rangle$ increases rapidly due 
to initial state scattering of the $c\bar c$ pair
\cite{Blaizot:1989hh};
this increase saturates when anomalous suppression sets in, 
and $\langle p_T^2\rangle$ eventually decreases above the knee 
of the $E_T$ distribution. 
Compared to the calculation in \cite{Kharzeev:1997ry}, this 
decrease is amplified by $E_T$ fluctuations. This
behavior is not compatible with the centrality dependence of 
$\langle p_T^2\rangle$ recently measured by NA50 \cite{Abreu:2000xe}. 
Indeed the data show a monotonous increase with $E_T$. 
Note however that our suppression criterion is
$p_T$-independent, while various arguments lead us to expect that a
$J/\psi$ with a high $p_T$  is to be less suppressed. The
implementation of such effects is under way \cite{progress}. 

\section{Predictions for RHIC}

The recent PHOBOS measurement \cite{Back:2000gw} shows that the 
total multiplicity is larger by 70\% 
at RHIC than at SPS in central collisions. 
We thus assume that the energy density is also increased by 70\% 
at RHIC. The above model then yields a parameter free prediction for
Au-Au collisions  at RHIC, which is plotted as the long-dashed curve
in Fig.\ref{f:plot}b. In order to compare RHIC and SPS results, we
have rescaled the RHIC tranverse energy so that the knees of the two
distributions coincide. 
In the Glauber model formulas, we use the value of the
nucleon-nucleon inelastic cross section $\sigma_{NN}=41$ mb at RHIC
energies, instead of $\sigma_{NN}=32$ mb at SPS energies. 

We have assumed so far that the energy density scales with the 
density of participants, i.e. 
$\epsilon \propto n_p(\mathbf b,\mathbf s) E_T/\langle E_T\rangle
(\mathbf b)$; this was a consequence of the observation that 
multiplicities and transverse energies approximately scale 
with the number of participants at SPS.
But at RHIC energies, a so-called ``hard'' component appears 
which gives rise to an additional term, proportional
to number of binary collisions \cite{Capella:2000zp,Kharzeev:2000ph}. 
The density per unit 
transverse area of binary collisions is 
$n_{\rm bin}(\mathbf b,\mathbf s) \propto T_A(\mathbf s) T_B(\mathbf {b-s})$,
where $T_A(\mathbf s)$ is the nucleus profile function, 
and this term must be taken into account in evaluating the 
energy density. 
We have checked numerically that this does not change qualitatively 
our predictions for $J/\psi$ suppression. 

Since the energy density is significantly larger at RHIC than at SPS, 
the anomalous suppression sets in earlier. Around the knee, the
suppression for Au--Au collisions is about 2.5 times greater than  at SPS for
Pb--Pb collisions, so that the effect of $E_T$ fluctuations, although
still visible, is much reduced. In a smaller system, e.g. Ca--Ca, the
$J/\psi$ suppression would be roughly the same at RHIC as the Pb--Pb system
at SPS. 
Note, however, that the present model ignores the possibility 
that several $c\bar c$ pairs may be produced in the collision. 
As recently shown \cite{Braun-Munzinger:2000px}, 
this may lead to an enhancement of $J/\psi$ production at RHIC, 
and may mask the suppression mechanism discussed here. 
\bigskip

\noindent{\bf Acknowledgments}
\medskip

We thank Bernard Chaurand for detailed explanations 
concerning the NA50 analysis.

\end{document}